\documentclass[noshowpacs,superscriptaddress,amsmath,amssymb,
  preprintnumbers]{elsarticle}
\usepackage{amsmath,amssymb,lmodern}
\usepackage{bbold}
\usepackage{bm}
\usepackage{relsize}
\usepackage{braket}
\usepackage{bigints}
\usepackage{graphicx}
\renewcommand{\tt}{\textit}

\renewcommand{\v}[0]{\bm}
\DeclareMathOperator{\Tr}{Tr}

\begin{document}
\title{Jet quenching test of the QCD matter created at RHIC and the LHC needs
  opacity-resummed medium induced radiation.}
\author[all]{Xabier Feal\corref{cor1}}
\ead{xabier.feal@igfae.usc.es}
\author[all]{Carlos A. Salgado}
\ead{carlos.salgado@usc.es}

\author[all]{Ricardo A. Vazquez}
\ead{vazquez@igfae.usc.es}

\address[all]{Instituto Galego de F\'{\i}sica de Altas
  Enerx\'{\i}as IGFAE, Universidade de Santiago de Compostela, 
   Galicia-Spain }
\cortext[cor1]{Corresponding author}
\date{\today}
\begin{abstract}
After almost two decades of investigation, jet quenching has become a
fundamental tool to study the properties of the QCD matter produced in
high-energy nuclear collisions.  Despite the large progress in both
experimental and theoretical tools, several unknowns remain to be
solved. In particular the systematics of the QCD matter parameters
extracted at different energies and centralities is still puzzling. On the other
hand, a lot of attention has been put in the last few years to a correct
resummation of the multiple scatterings that the particles of the jet suffer
when traversing the QCD matter. Present formalisms have substantially 
improved the multiple-soft or single hard scattering approximations, common in 
the last twenty years. We show here that medium parameters extracted with
 full resummed medium-induced gluon radiation spectra significantly differ
 from these two extreme approximations, providing 
 a natural systematics for different 
 collision energies and centralities, with no inconsistencies.  For the 
 case of the full resummation this ensures that the density of scattering
 centers as a function of the temperature qualitatively agree with the
  equation of state computed in 
 lattice QCD. 
\end{abstract}

\maketitle
The hottest matter ever created in laboratory, the quark gluon plasma (QGP),
is produced by smashing large atomic nuclei in particle colliders. The QGP is
thought to be the primordial material that filled the whole universe some
micro-seconds after the Big Bang and measuring its properties is the main goal
of the experimental programs of nuclear collisions at RHIC and LHC.  These
{\it heavy ion} programs have delivered an outstanding body of experimental
evidence and novel theoretical developments supporting our current view on the
QGP as a nearly inviscid, rapidly thermalized liquid, color screened to
melt any hadronic bound state. Determining the properties of this phase of the
matter has constituted one of the most important open problems in high energy
physics and a milestone in our understanding of the theory of the strong
interactions \cite{akiba2015,Citron,Dainese}.

One of the main experimental tools in these studies is jet quenching:
fast particles retain characteristic imprints of their passage through
the QGP that can be measured as modifications in the high transverse
momentum spectrum of the collision.  A good theoretical control on the
underlying dynamics is a prerequisite to relate these measured
imprints with the properties of the QGP. At high enough energy, the
dynamics is dominated by induced radiation of gluons. In simple
quantum-mechanical terms, this radiation can be understood as the
decoherence of a gluon fluctuation from the fast particle due to
color-rotating multiple elastic collisions with the medium. The
elementary cross section of these elastic collisions has been computed
in perturbative QCD. The medium-induced gluon radiation spectrum has
also been known for some time in the soft limit
\cite{Zakharov:1996fv,baier1997a,gyulassy2000,wiedemann2000b} and
beyond the soft approximation \cite{apolinario2014,sievert2018}; see
also \cite{Blaizot:2012fh,Zakharov:1998sv}. In all cases, expressions
for an arbitrary number of scattering centers are known --- the {\it
  opacity expansion} series --- but difficult to treat
numerically. For this reason, two limiting cases are usually employed:
i) keeping only the first term in the expansion but the full
perturbative cross section (also known as $N=1$ or single hard
approximation in the following, sometimes also referred as GLV
approximation); ii) resumming all multiple scatterings in Gaussian
approximation, that neglects the perturbative power-law Coulomb tails
(also known as BDMPS or multiple soft scattering approximation). While
these two approximations have provided a good description of the
suppression for a given colliding energy and centrality, a common
understanding of RHIC and LHC data at different centralities and
energies has been puzzling our community for almost 10 years. An
inconsistent deviation by a factor $K\simeq 1.3-2$ in the extracted
QGP opacity has been systematically found in all the phenomenological
analyses to date when going from LHC to RHIC energies. Triggered by
this unsatisfactory understanding of more and more precise data on jet
quenching, the relevance of the resummation schemes have recently
received a renewed attention (see
e.g. \cite{Mehtar-Tani:2019tvy,Sievert:2019cwq,Andres:2020vxs,Barata:2020sav}). To
this extent, it is natural to wonder whether the absence of a correct
resummation may have been reabsorbed in the extracted opacity
parameter as a systematic deviation.  The main goal of this letter is
to show that this is indeed the case, i.e. that a full resummation of
scattering centers including the expected perturbative tails, modifies
the systematics of the extracted medium parameters with collision
energy and centrality, providing a natural solution to the
above-mentioned puzzle. We find that the extracted number density (the
fitting parameter in our approach) qualitatively agrees with the QGP
equation of state computed in lattice QCD, while it qualitatively
disagrees if the resummation is not performed. In most
phenomenological studies, the QCD equation of state is an input of the
calculation and the disagreement is absorbed in modified medium
parameters ---often through the strong coupling constant
$\alpha_s$. Our different strategy makes the origin of the
disagreements evident, shows that the full resummation is needed for a
precise determination of the medium parameters and provides a
framework to determine the QGP color opacity with jet observables
with no inconsistencies. 

In this exploratory study, we shall not attempt to model the precise
medium dynamics but adopt a simpler approach in which once the 
temperature at some initial time is fixed, all other quantities in the calculation,
except the number density, are computed from known perturbative 
relations. The number density is then the free parameter in the fit of the 
inclusive particle suppression $R_{AA}$ for each 
collision energy and centrality. Using this procedure, somehow orthogonal to the
usual implementation in which the density is directly taken from e.g.
hydrodynamical profiles, we
can check the temperature
dependence of the number density and perform the qualitative comparisons\footnote{We 
work under the assumption that the QGP admits a
quasiparticle interpretation for which our perturbative description
would be adequate -- see also \cite{DEramo:2012uzl}. For an ideal gas the number density
can be directly related to other thermodynamical quantities, but this is not
the case for an interacting QGP and this is why our comparison 
has to be taken at the qualitative level at this stage.}
 with available 
results from lattice QCD on the equation of state that we have mentioned above. 

%
To evaluate
the gluon bremsstrahlung we will follow \cite{feal2018a} and
discretize the gluon path in $n$ steps of length $\delta z=
0.1\lambda_g$, where $\lambda_g$ is the gluon mean free path. The gluon
four momenta is given by $k^l=(\omega,\v{k}^l)$ for
$l=1,\ldots,n$. Soft gluons in this picture can be emitted from any of
the $l$-th internal lines or from the last $n$-th leg. These two
classes of diagrams can be accounted for with color-factorized
4-currents $J^l$ and $J^n$, respectively, whose components in the
Coulomb gauge read
\begin{align}
  J^l_i= \epsilon_{ijk}\frac{ k_j^lp_k}{k_\mu^l p^\mu}
  \left(e^{i\varphi_{l+1}^n}-e^{i\varphi_{l}^n}\right),
  \medspace\medspace\medspace
  J^n_i= \epsilon_{ijk}\frac{ k_j^np_k}{k_\mu^n p^\mu},
  \label{eq:currents}
\end{align}
and $J_0^l=J_0^n=0$. Here, the hard particle 4-momentum $p$ has been
left fixed along the initial direction and $\epsilon_{ijk}$ is the
Levi-Civita symbol. The decoherence of phases $\varphi^{a}_{b}$
between gluon interactions at $a$ and $b$ modulates the non-Abelian
LPM suppression \cite{landau1953a,landau1953b,migdal1956}, and reads
\begin{align}
  \varphi^n_l=\frac{1}{p_0}\sum^{n-1}_{i=l}\delta z_i k^i_\mu p^\mu,
  \medspace\medspace\medspace
  \delta z_i = z_{i+1}-z_i,
\end{align}
where $z_i$ is the path discretization along the traveling direction and $p_0$
the hard particle energy. The total intensity is measured with respect to the
(zeroth order) radiation at the hard production vertex
\cite{baier1997a,baier1997b}, i.e. the radiation still present in the absence
of a medium,
\begin{align}
J_T^2(k)=\bigg|J^n+\sum_{l=1}^{n-1}
J^l\bigg|^2-\bigg|J^n\bigg|^2,
\label{total_current}
\end{align}
and has to be evaluated within the internal gluon momentum distributions. In a
tagged parton scenario in which color and spatial averages can be taken, event
by event fluctuations fade out and the gluon in-medium dynamics can be
stochastically built with the knowledge of the color averaged cross section of
changing momentum $\v{q}$ into a single collision, whose Fourier transform
reads
\begin{align}
  \sigma(\v{x})\equiv \int \frac{d^2\v{q}}{(2\pi)^2}e^{-i\v{q}\cdot\v{x}}
  \frac{1}{d_Ad_R}\Tr
  \left(F_{el}^{\dag}(\v{q})F_{el}(\v{q})\right),
  \label{single_elastic_distribution}
\end{align}
with $d_A=N_c^2-1$ and $d_R=N_c$ ($d_R=N_c^2-1$) the color dimension
of the target quark (gluon). For the relevant temperatures of the hot
medium the single scattering amplitudes can be expanded at leading
order in the coupling as $F_{el}(\v{q})=-ig_s^2t_\alpha^A
t_\alpha^R/(\v{q}^2+\mu_d^2)$ with $g_s^2=4\pi\alpha_s$, where $t_A$
and $t_R$ are SU(3) matrices in the gluon and the target
representation and $\mu_d= 1/r_d$ the screening mass of the color
field of a single scattering center. After an arbitrary number of
collisions governed by \eqref{single_elastic_distribution}, the
probability of emerging with momentum change $\v{q}$ in a medium path
$s$ is given by
\begin{align}
  \phi(q,s) = 2\pi\delta(q^0) \int d^2\v{x}e^{i\v{q}\cdot\v{x}}\exp\int^s_0 dz
  \rho(z)\bigg[\sigma(\v{x}) -\sigma(\v{0})\bigg],
  \label{multiple_elastic_distribution}
\end{align}
where $\rho(z)$ is the local number density of the medium at a depth $z$. This
distribution satisfies the Moliere QCD equation with kernel
\eqref{single_elastic_distribution} and inherits, though with larger widths
linear with $l$, the long $\v{q}$ tails of the single scattering scenario. The
intensity of soft gluons emitted in an in-medium path $l$ in the energy
interval $\omega$ and $\omega+d\omega$ per unit of time and per unit of medium
transverse size satisfies the weighted convolution of \eqref{total_current}
with \eqref{multiple_elastic_distribution}
\begin{align}
\omega \frac{dI}{d\omega} = \alpha_s C_R&
\int \frac{d\Omega_n}{(2\pi)^2}\left(\prod_{k=0}^{n-1}\int 
\frac{d^3\v{k}_{k}}{(2\pi)^3}\phi(q_k,
\delta z_k) \right)J_T^2(k),
\label{squared_emission_amplitude}
\end{align}
with $\Omega_n$ the gluon solid angle, $C_R=4/3\medspace (3)$ the
color averaged charge of the squared vertex $q\to qg$ ($g\to gg$) and
$q_k=k_{k+1}-k_{k}$. For consistency within the soft gluon approximation the splitting
function corrections \cite{zakharov1996,sievert2018} are here
omitted\footnote{We have checked that for the most central \textit{PbPb} collisions
($\sqrt{s_{NN}}$=2.76 TeV) this results in medium density
underestimations of $\sim 22\%$.}. The expression given above
Eq.\eqref{squared_emission_amplitude} can be interpreted in
probabilistic terms and is suitable for a Monte Carlo evaluation by
summing over many gluon in-medium paths of length $l$. In the
continuous limit path integrals on the light cone are recovered
\cite{feal2018b}. In addition, evaluation of
Eq.\eqref{squared_emission_amplitude} in an expanding medium is
straightforward. Closed results, like the GLV
\cite{gyulassy2000,wicks2008} and BDMPS-ASW
\cite{baier1997a,baier1997b,wiedemann2000,salgado2003,armesto2004}
predictions for small and arbitrary opacities, respectively, can be
obtained from Eq.\eqref{squared_emission_amplitude} either with
perturbative expansions or by assuming Brownian elastic scattering
respectively
\cite{feal2018b}. 
Multiple hard interactions in semi-infinte media
have been also resummed before not exactly considering the
radiation scenario after the hard production vertex \cite{arnold2002}
or making use of the Gaussian approximation \cite{zakharov1996}, and
then later extended for the finite case
\cite{CaronHuot:2010bp}. Radiation \eqref{squared_emission_amplitude}
has a typical width characterized by the scale $\omega_c\simeq
1/2\hat{q}l^2$ \cite{baier1997a,baier1997b,salgado2003}, where
$\hat{q}$ is the QGP transport coefficient.  Gluons of energy
$\omega>\omega_c$ are not able to resolve the medium \cite{feal2018a}
and thus radiation vanishes slowly with large tails $\propto
1/\omega$. At the smallest energies $\omega\leq \mu_d$ kinematical
constraints also cancel the radiation.

\begin{figure}[ht]
  \begin{centering}
    \includegraphics[scale=0.65]{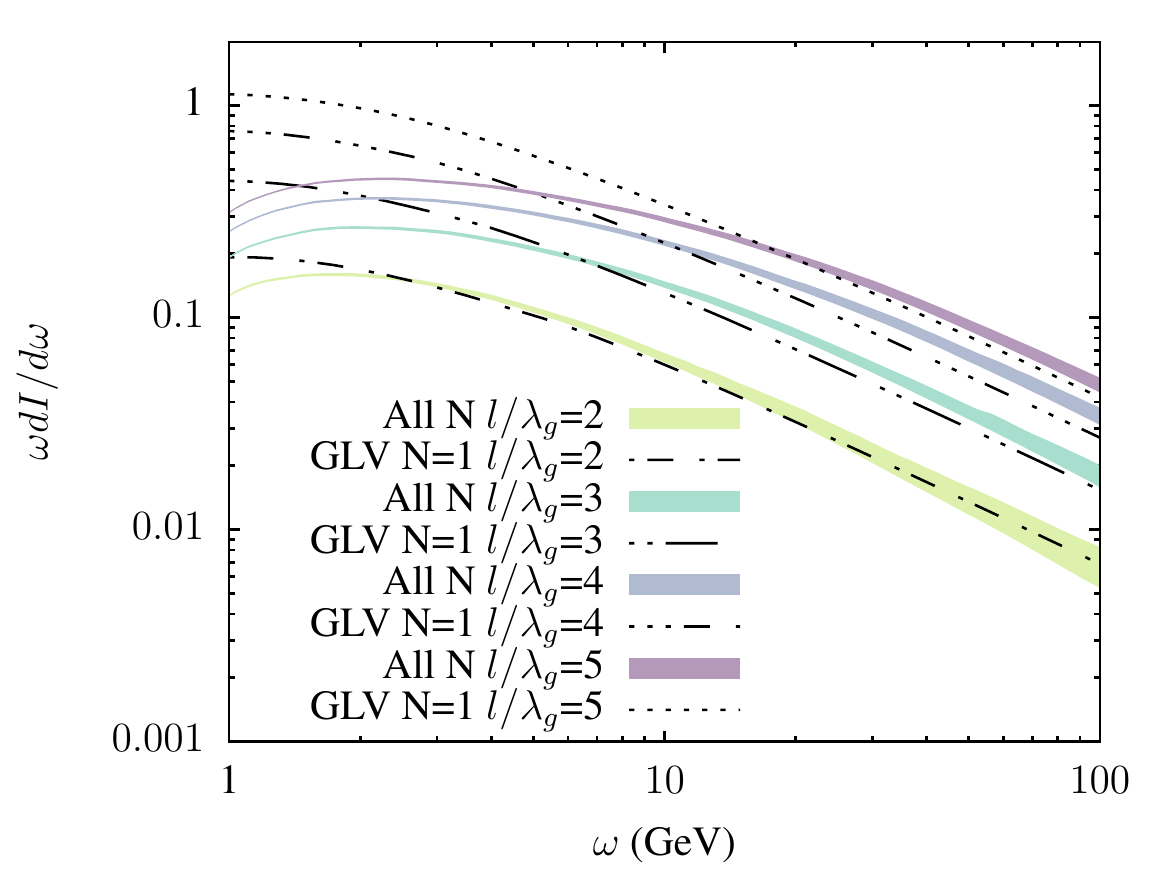}
    \caption{Medium induced gluon radiation for a QGP brick involving $l/\lambda_g$=2,3,4 collisions on average, in the N=1 term in an opacity expansion (GLV) and the full resummation. }
    \label{fig:comparison}
  \end{centering}
\end{figure}

Our numerical implementation of the formalism described in eqs. 
(\ref{eq:currents})-(\ref{squared_emission_amplitude}) allows to perform
the exact resummation of the multiple scattering contributions with realistic 
scattering potentials -- we have used a Debye potential but very similar
results can be obtained for thermal potential \cite{Andres:2020vxs}. 
In Fig. \ref{fig:comparison}
we present our full results  compared with the
case in which only a single scattering is considered for different opacity parameters
of relevance for the present phenomenological study. 
One clearly sees that increasing the average number of scattering centers
makes the spectrum to depart more and more from the single scattering
case which has the tendency to overpopulate the soft gluon radiation.
As a consequence, when the density of the medium grows (e.g. when going 
from lower to higher collision energies) overestimating the soft gluon
radiation translates into an artificially larger energy loss that can only be compensated
by the corresponding reduction of the  medium parameters in order to reproduce the
experimental data\footnote{In fact, the actual effect is a bit more complicated as it
involves the different slopes in the perturbative spectrum convoluted with the energy loss
probabilities. These slopes are smaller at larger energies --- 
the spectra of produced particles is harder --- making it more sentitive to the infrared.}. 
We identify this as the main effect in the energy and centrality
puzzle for the systematics of the medium parameters.

We should notice that under the intensity \eqref{squared_emission_amplitude}
more than a single gluon is usually emitted. For soft gluon emission, when
the formation time is much smaller than the length of the medium, the daughter
partons in the splitting lose color coherence very fast and an independent 
gluon emission is justified \cite{Blaizot:2012fh, apolinario2014}.
Here, we take the usual Poisson approximation for the 
probability $P(\epsilon,p_t)$ of a hard
parton of energy $p_t$ to lose an energy $\epsilon$ ---also known as
the quenching weight
\cite{baier2001a,salgado2003},
\begin{align}
  P(\epsilon,p_t)=\mathcal{N}\sum_{n=0}^\infty \frac{1}{n!}
  \delta\bigg(\epsilon-\sum_{i=1}^n\omega_i\bigg)\prod_{i=1}^n
  I\bigg(p_t-\sum_{j=1}^{i-1}\omega_j\bigg).
  \label{quenching_weight}
\end{align}
where the average number of gluons,
$I(p_t)$, and the probability of no emission $N(p_t)$, are functions of the
initial energy 
\begin{align}
  I(p_t)=\int^{p_t}_{0}d\omega\frac{dI}{d\omega},
  \medspace\medspace\medspace\medspace\medspace\medspace
  N(p_t)=e^{-I(p_t)}.
\end{align}
%


%
In this work, we are interested in the suppression of inclusive particle
production at large transverse momentum, where the
 detected particle is not  a parton $i$, but a given hadron,
$h$, carrying away a fraction $z$ of the parton energy with probability
$D_i^h(z)$. The abundance of hard events may be assumed to scale with the
number of hard processes in a single nucleon-nucleon collision
\cite{miller2007} and within these pQCD assumptions the quenched inclusive
hadron cross section reads \cite{baier2001a,arleo2017}
\begin{align}
  \frac{d\sigma_{AA}^h(p_t)}{dydp_t}=& T_{AA} \sum_i\int dz
  D_{i}^h(z)\int^\infty_0 d\epsilon
  \frac{d\sigma_{pp}^i(p_t/z+\epsilon)}{dy dp_t}
  P_i(\epsilon,p_t/z+\epsilon),
\end{align}
where $T_{AA}$ is the nuclear overlap function. In order to make the
numerical computations treatable, in this first study we consider
$P_i(\epsilon,p_t/z+\epsilon)$ to be smooth enough to allow an
evaluation at the typical fragmentation ratio $\langle z \rangle$ to
find \cite{salgado2003,baier2001a}
\begin{align}
  \frac{d\sigma_{AA}^h(p_t)}{dydp_t}\simeq T_{AA}\sum_{i}\int^{\infty}_0 d\epsilon
  \frac{d\sigma_{pp}^{i,h}(p_t+\epsilon)}{dy dp_t}P_i(\epsilon,p_t+\epsilon).
  \label{hadron_cross_section}
\end{align}
with $i$ running over gluons and the different quark species. This
approximation has been frequently used (see
e.g. \cite{salgado2003,baier2001a,arleo2017}) as it provides a quite
precise answer with a much simpler numerical implementation, as
discussed also in \cite{Eskola:2004cr}. Direct calculations of the 
sensitivity of the hadron spectrum suppression to different 
fragmentation function scenarios is also known to be small ---see
e.g. \cite{Andres:2019eus}. Hadron fragmentation, typically to pions,
takes place via the gluon channels at small $p_t$ and it is dominated
by quarks starting from $p_t$ around 30 GeV to 100 GeV for
$\sqrt{s_{nn}}=$0.9 TeV to 7 TeV, depending on the different
fragmentation implementations
\cite{stratmann2010,denterria2014}. Since gluons lose noticeably more
energy than quarks due to their larger color space factor in
\eqref{single_elastic_distribution} and
\eqref{squared_emission_amplitude}, 
the suppression of high-$p_t$ hadrons will be larger when they originate
from gluons than for quarks. In order to implement this difference we
take the gluon-to-quark ratio in the hadronic spectrum to be 
\begin{align}
  \frac{d\sigma_{pp}^{g,h}(p_t)}{dydp_t}\bigg/
  \sum_i \frac{d\sigma_{pp}^{i,h}(p_t)}{dydp_t}=1-\frac{1}{2p_t^c}p_t,
  \label{eq:qtog}
\end{align}
with the quark/gluon turnover around $p_t^c\simeq$ 60 GeV fixed for
all the collision energies explored in this work. In this simplified 
treatment of the perturbative spectrum we just classify the parent
partons in gluons or quarks, irrespectively of the flavor, so 
the sum in the denominator of Eq. (\ref{eq:qtog}) 
runs over all quark flavors. This setup fulfills
the single gluon (quark) channel scenario at $p_t\simeq 0$ ($p_t\simeq
2p_t^c$), respectively, and deviates only a $\sim 10\%$ from some
fragmentation implementations \cite{denterria2014}. We have 
checked that this deviation translates into QGP density
predictions affected by $\sim 8\%$ at the most central \textit{PbPb}
collisions ($\sqrt{s_{nn}}$=2.76 TeV). As usual, 
the nuclear modification factor
is given by
\begin{align}
R_{AA}^{h}\equiv  \frac{1}{\langle T_{AA}\rangle
}\frac{d\sigma_{AA}^{h}(p_t)}{dydp_t}\bigg/
\frac{d\sigma_{pp}^{h}(p_t)}{dydp_t}.
\label{nuclear_modification_factor}
\end{align}
Finally, the $pp$ inclusive hadronic cross section in
\eqref{nuclear_modification_factor} is taken in this analysis directly
from the experimental data, so that no further uncertainty is needed.

The inputs to the spectrum
\eqref{squared_emission_amplitude}, that completely determines
the energy loss distribution \eqref{quenching_weight} and, hence, 
the suppression  \eqref{nuclear_modification_factor},
are the Debye screening mass $\mu_D$, the running coupling 
$\alpha_s$, the medium length traversed by the parton $l$ and
the density of scattering centers $\rho$, that is the only fitting 
parameter in our approach. We discuss now how we fix
these different inputs.

After formation the QGP thermalizes at some temperature
$T_0$ in a time $\tau_0\lesssim$ 1 fm/c following non-equilibrium
dynamics.  We will simply assume that thermalization is achieved
quicker when the number of collisions is larger $\tau_0\propto 1/T_0$
\cite{baym1984,baier2008}. The system subsequently expands and cools
down with lifetimes of the order $\lesssim$ 10 fm/c according to event
by event dynamics that can be only accessed through precise
hydrodynamical or kinetic theory calculations (see
e.g. \cite{strickland2018} for a recent review). The fast parton
quenching, however, is mostly sensitive to the microscopic scales
$\mu_d$ and the path-averaged macroscopic opacity in $\omega_c$
\cite{salgado2002,djordjevic2008}, i.e. to the average number of
collisions along the path of its travel. 
It has become customary to obtain these local properties from 
 hydrodynamical studies of the heavy-ion collision data
in the soft part of the spectrum. 
Here, however, this usual prescription would prevent us
to check the (qualitative) agreement of the temperature dependence 
of the number density with the equation of state. For this reason, we
assume  a simple one-dimensional longitudinal expansion
\cite{bjorken1983,baier1998} $\rho(\tau)=\rho(\tau_0)\tau_0/\tau$ and
accurate estimations of the QGP lifetimes through Bose-Einstein pion
correlations \cite{makhlin1988,retiere2004}. Several works have
shown that the experimental accuracy of $R_{AA}$ is not enough to 
distinguish between different hydrodynamical profiles  ---see e.g. 
\cite{andres2016}.

In order to fix the length that the parton travels through matter, we 
consider that the
typical lifetimes of the QGP are usually smaller than its initial
transverse dimensions, varying in the most central collisions from
$\tau_f$=7.3$\pm$0.2 fm/c in \textit{AuAu} collisions
($\sqrt{s_{nn}}$= 200 GeV) to around $\tau_f$=9.8$\pm$0.9 fm/c in
\textit{PbPb} collisions ($\sqrt{s_{nn}}=$ 2.76 TeV). The combined
data can be parameterized into a single expression
$\tau_f$=(0.87$\pm$0.01)$\times(dN_{ch}/d\eta)^{1/3}$ fm/c, deviating
less than a 10$\%$ from the RHIC and LHC predictions with freeze-out
at $T_f\simeq$ 120 MeV \cite{star2005b,alice2015}. We will only
attempt here to evaluate \eqref{nuclear_modification_factor} for the
averaged in-medium path length constrained by these HBT system sizes.
For typical $p_t\gtrsim$ 5 GeV the interactions with the cold nuclear
matter can be neglected \cite{alice2018a}, then on average the fast
parton travels a distance
\begin{align}
l\simeq \int_0^{\tau_f} p(s) sds+\tau_f\int_{\tau_f}^{2R} p(s)ds,
\end{align}
subject to the hot medium quenching, where for simplicity and within
our current uncertainties isotropic production may be assumed
$p(s)=\sqrt{4R^2-s^2}/(R^2\pi)$ and $R$ is the radius of the nuclear
overlapping area. A more precise implementation of the path
fluctuations and the hard production anisotropies are required for a
better treatment of the low $p_t$ regime and the most peripheral
collisions. Typical in-medium path lengths are found at the largest
RHIC energies 10-20$\%$ smaller than their LHC counterparts. The
effective volume found by the parton within this setup is in very good
agreement, in the analyzed range of centralities, with the
hydrodynamical predictions at the freeze-out \cite{gardim2019}.

\begin{figure}[ht]
\centering
\includegraphics[scale=0.65]{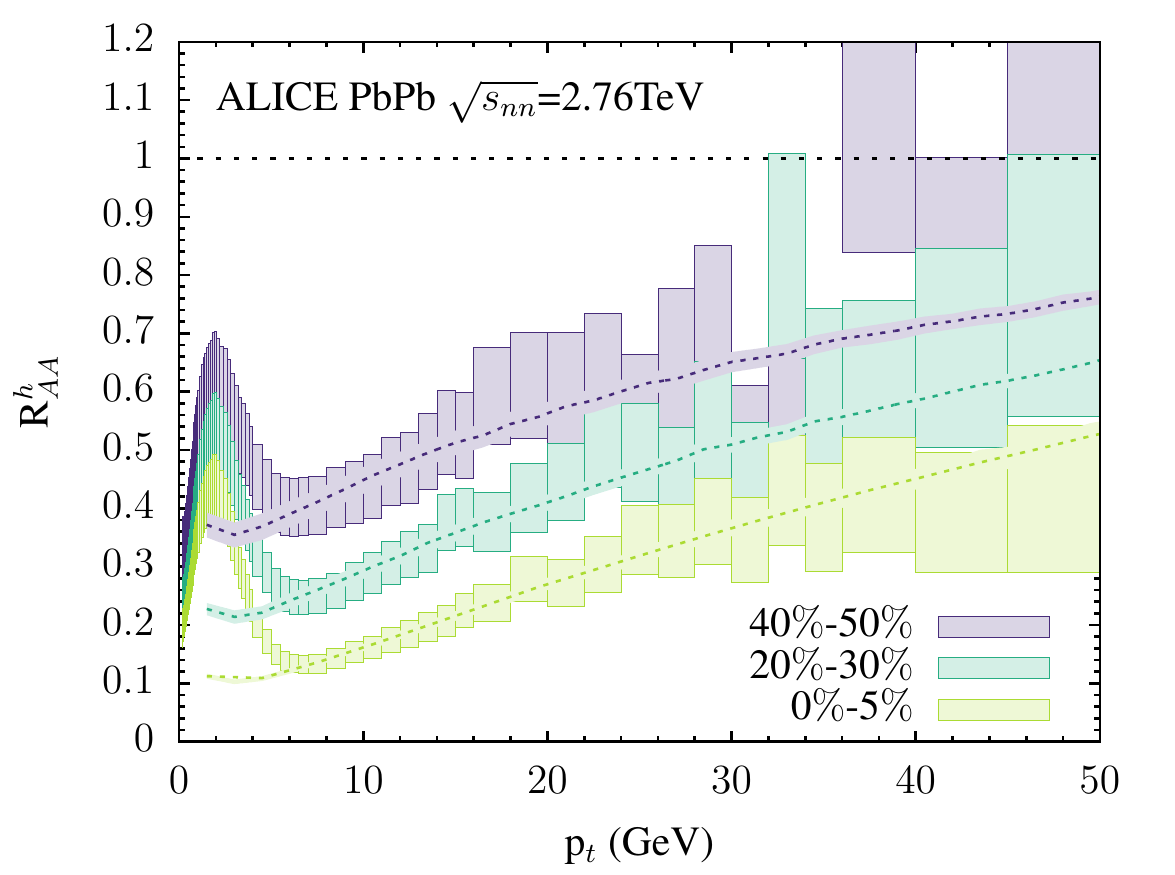}
\caption{Nuclear modification factor $R_{AA}^h$ as a function of
  transverse energy $p_t$ for a QGP with initial density at the most
  central class $\rho_0$$\simeq$56/fm$^3$, $\alpha_s$$\simeq$0.28,
  $\mu_d$$\simeq$ 1.1 GeV, $\tau_0$$\simeq$0.6 fm/c and lifetime
  $\tau_f$$\simeq$9.8 fm/c fitted to ALICE data \cite{alice2013a}.}
\label{figure_1}
\end{figure}
The running coupling is set at leading order
$\alpha_s(q)=1/(b_0\ln(q^2/\Lambda^2))$, with $b_0=(33-2N_f)/12\pi$
the 1-loop $\beta$-function coefficient for $N_f=2+1$ active flavors.
The latest world average at the Z$^0$ mass
$\alpha_s(M_Z^2)$=0.1181$\pm $0.0011 fixes the QCD scale at
$\Lambda_s=247$ MeV \cite{pdb2018}. This setup reproduces well the
collected data below $\alpha_s(m_\tau)$=0.325$\pm$0.015 and then it
shall be considered safe for the collision energies explored in this
work. The screening length of the QGP may be set with the help of the
HTL \cite{braaten1990,rebhan1993} result $\mu_d^2(T)=4\pi\alpha_s(2\pi
T)(1+N_f/6) T^2$ which within the above coupling setup and the running
scale set to $q=2\pi T$ matches the heavy-quark free energy
predictions in quenched lattice computations ($N_f=0$) and falls a
10$\%$ below the full QCD result with two flavors in the staggered
quark action \cite{zantow2005}.

We have made an analysis of existing data on jet quenching, including
\textit{CuCu} and \textit{AuAu} data at 200 GeV
\cite{phenix2008a,phenix2008b}, \textit{PbPb} at 2.76 TeV
\cite{cms2012,alice2013a,alice2011}, \textit{PbPb} at 5.02 TeV
\cite{cms2017,alice2016c}, and \textit{XeXe} data at 5.44 TeV
\cite{cms2018}. For each centrality and energy considered a fit is
made to the nuclear modification factor $R_{AA}^h$ using $\rho_0$ as
the single unknown parameter.  To determine the 
initial
temperature of each
analyzed collision system, energy and centrality we used $\epsilon
\tau_0\propto T_0^3$ measurements, when available, and extrapolated
the relation $\epsilon\tau_0\simeq (8.85 \pm 0.44) \times
(\sqrt{s_{nn}})^{0.33\pm 0.02}$ GeV$^2$/fm for the most central
collisions between $\sqrt{s_{nn}}=$ 27 GeV-2.76 TeV, when measurements
have not yet been made available \cite{phenix2016,alice2016b}. We then
fix as a reference the most central \textit{PbPb} collisions at
$\sqrt{s_{nn}}=2.76$ TeV to a temperature of $T_0\simeq$ 470 MeV
\cite{kolb2003} and $\tau_0$=0.6 fm. This temperature then fixes with
a single setup all the parameters in the analysis, whose 
temperature dependence was explained in the last paragraphs, except
$\rho_0$, which is taken as the free parameter for each centrality,
energy and collision system.

As a first example, fitting the most central \textit{PbPb} collisions
at $\sqrt{s_{nn}}=2.76$ TeV yields $\rho(\tau_0)\simeq 56$
fm$^{-3}$. The fit for this example case is shown in Fig.\ref{figure_1}, where
we plot $R_{AA}^h$ as a function of $p_t$ for three centrality classes
and include in the caption the numerical values of the QGP parameters
obtained for the most central data. The initial density is found to
scale roughly proportional to $N_{part}^{1/2}\propto T_0^3$ at fixed
collision energy. At the largest RHIC energies $\sqrt{s_{nn}}=$ 200
GeV in the most central \textit{AuAu} collisions the initial
temperature extracted from the energy density measurements yields
$T_0\simeq 362$ MeV and the density obtained in the fit
$\rho(\tau_0)\simeq 21$ fm$^{-3}$.

\begin{figure}[ht]
  \centering
  \includegraphics[scale=0.65]{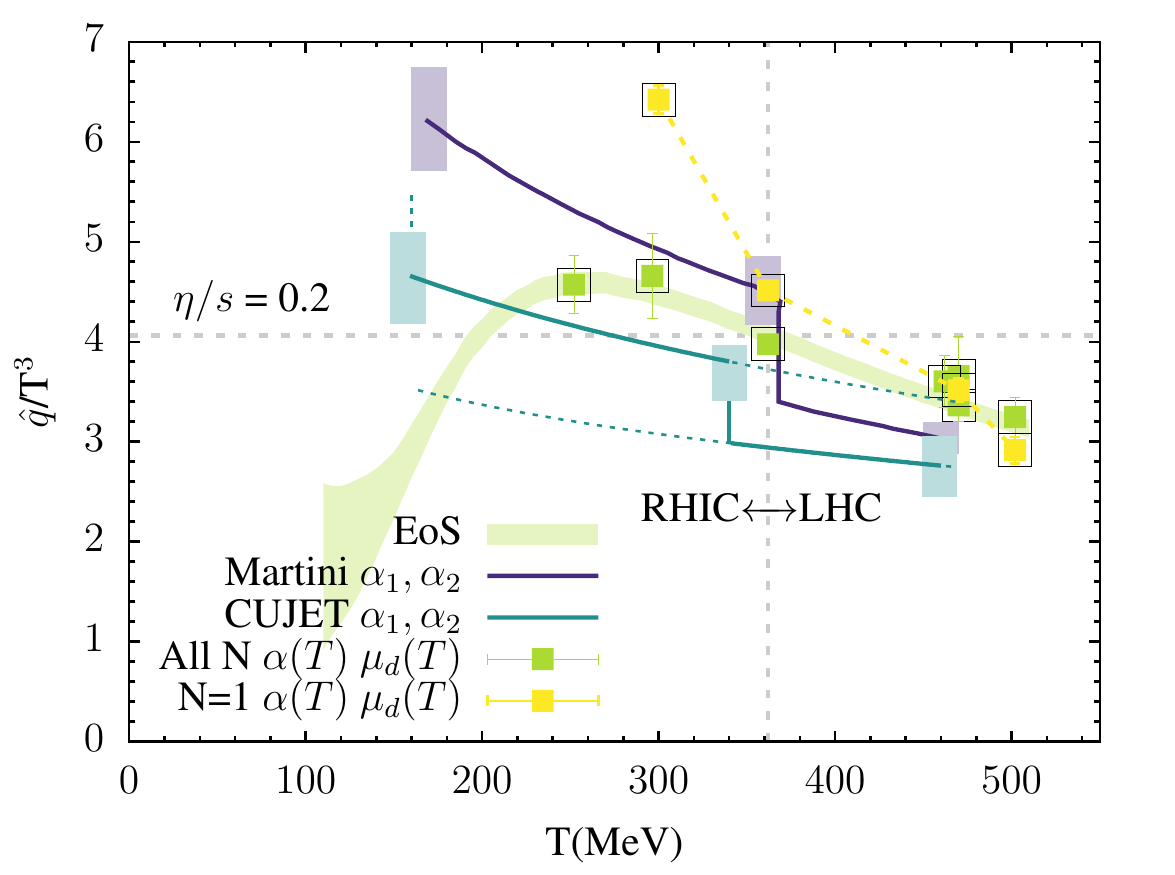}
  \caption{QGP transport parameter $\hat{q}$ for a gluon of
    $\omega$=10 GeV, using the density extracted from an all order
    (green squares) or a fist order (yellow squares) jet quenching
    analysis of same data as Fig. \ref{figure_2}. Also shown is the
    $\hat{q}$ assuming $\rho=p/T^4$ from lattice predictions of the QCD
    Equation of State \cite{wuppertal2016} (green band), and the CUJET
    (blue) and MARTINI (purple) puzzles found in \cite{jet2014}.}
  \label{figure_2}
\end{figure}

Our results on the fitting parameter $\rho$ scale roughly constant
with $T^3$, in agreement with expectations. 
The same analysis using the single hard approximation to
the gluon radiation, produces instead progressive deviations with
centrality in the scale $\rho/T^3$. A deviation factor of $K=1.22$ in
$\rho/T^3$ is found between the hottest QGP created at RHIC and the
most central collisions of \textit{PbPb} at $\sqrt{s_{nn}}=2.76$ TeV
by using the single hard approximation. With the coupling and tail
logarithmic corrections, this translates into a deviation of $K=1.29$
in $\hat{q}$,
\begin{align}
\hat{q}=2\pi\alpha_s^2C_R\rho_0\left(\log\left(\frac{2p_0}{\mu_d}\right)-\frac{1}{2}\right)
\end{align}
the QGP transport parameter. This finding is consistent
with the value $K\simeq 1.3$ found by previous perturbative analyses
\cite{jet2014}. These results for $\hat{q}$ are shown in
Fig.\ref{figure_2} and compared with the energy puzzles found by
\cite{jet2014}. While the single hard approximation ($N=1$) provides a
reasonable estimation of the $1/\omega$ tails of the radiation for the
RHIC and LHC scenarios - involving few collisions in general
$n_c\lesssim 5$ -, we have checked that large uncertainties, related
to the neglect of the soft resummations and the non-collinear effects,
are behind these energy puzzles. Similar deviations, however with
larger factors $K\simeq 2$, have been also found by the Gaussian
resummations of the gluon spectrum \cite{andres2016}, where the hard
tails are instead neglected. We then argue that, for a consistent
description of the color opacity and $\hat{q}$ without temperature
issues, an accurate implementation of the underlying quenching
mechanisms is necessary beyond the single hard (GLV) and the multiple
soft (BDMPS) approximations.

\begin{figure}[ht]
  \centering
  \includegraphics[scale=0.65]{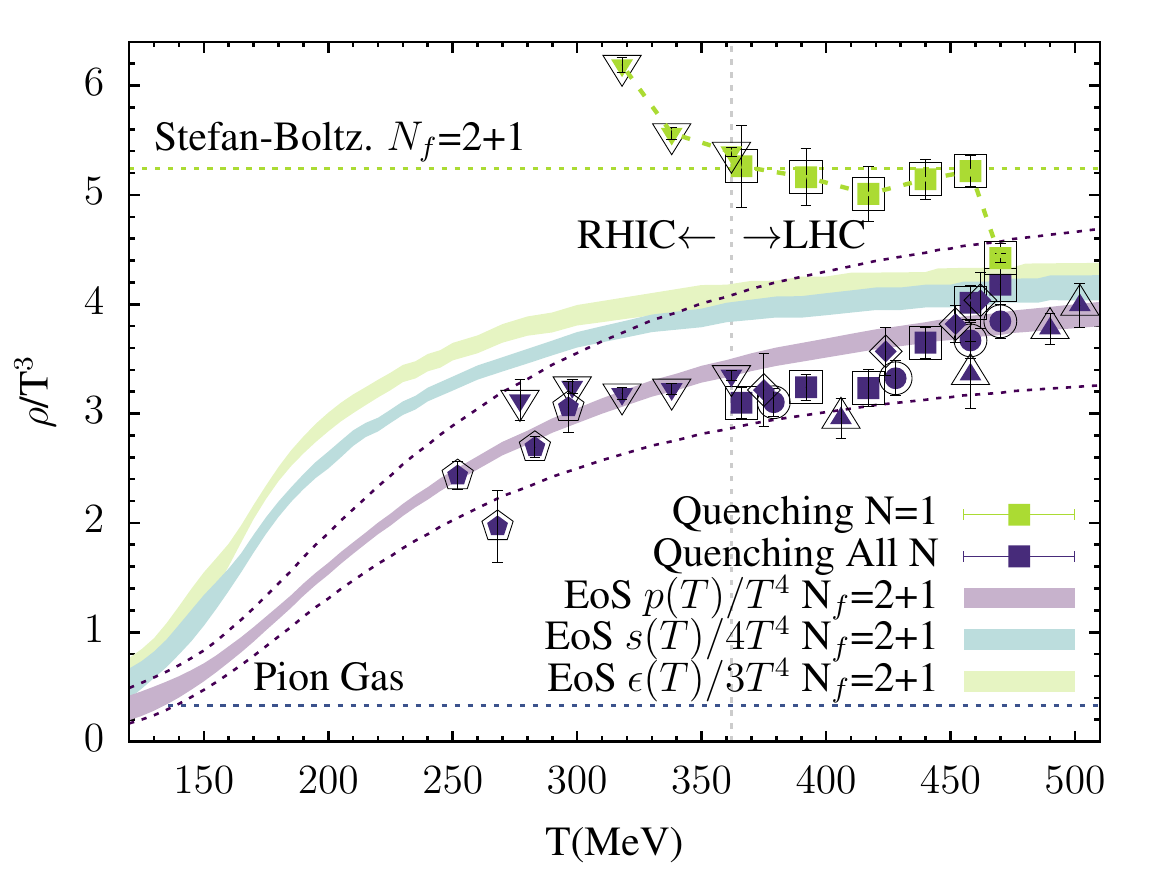}
  \caption{QGP density extracted from All $N$ (purple symbols ) and $N=1$
    (green symbols) analyses of the $R_{AA}^h$ collected data on collisions of
    \textit{CuCu} (pentagons) and \textit{AuAu} (down triangles) at
    $\sqrt{s_{nn}}=$200 GeV, \textit{PbPb} at $\sqrt{s_{nn}}$=2.76 TeV
    (squares and circles), \textit{PbPb} at $\sqrt{s_{nn}}$=5.02 TeV 
    (up triangles) and \textit{XeXe} at $\sqrt{s_{nn}}$=5.44 TeV
    (diamonds) from PHENIX, ALICE and CMS Collaborations, compared to
    lattice results of the Equation of State by the Wuppertal
    collaboration \cite{wuppertal2016}.}
  \label{figure_3}
\end{figure}

The ratio of the fitting parameter $\rho(\tau_0)$ with the initial
temperature $T_0$ is plotted in Fig.\ref{figure_3} for all the
different collision systems, energies and centralities considered.
Also shown is the lattice result of $p/T^4$, $s/(4 T^3)$, and
$\epsilon/(3T^4)$ with $N_f= 2+1$ \cite{wuppertal2016}.  A global
uncertainty of the $17\%$ --- $28\%$ if fragmentation and splitting
function effects were considered --- is marked with bands, indicating
a $10\%$ of variation of all the unknown parameters in our study.  The
trend of our results agree well, without temperature issues with the
rescaled QCD equation of state. For an ideal relativistic gas, it is
verified that $\rho/T^3= p/T^4 = s/(4 T^3)= \epsilon/(3 T^4)$,
however, for an interacting quark gluon plasma, this is not the case 
and the agreement should be taken at the qualitative level.
Our results provide then an interesting phenomenological approach to
study the high temperature behavior of the color opacity of the QGP,
as seen by the fast traveling partons in the collision. Collisional
energy losses \cite{peigne2008} are also neglected here and may also
affect the low $p_t$ fit. Although further studies are required, this
result indicates that accurate resummations of the medium-induced
gluon radiation, like the one presented in this letter, do solve the
jet quenching energy puzzles between RHIC and the LHC and may open up
an additional handle on the study of the QCD equation of state, for which
a more rigorous connection between the jet quenching parameters and
the QGP thermodynamical quantities would be needed.

\vglue 5mm
We thank useful discussions with Carlota Andr\'es, N\'estor Armesto,
and P\'\i a Zurita. This work has been funded by Ministerio de Ciencia e
Innovaci\'on of Spain under project FPA2017-83814-P; Unidad de Excelencia
Mar\'\i a de Maetzu under project MDM-2016-0692; ERC-2018-ADG-835105 YoctoLHC;
and Xunta de Galicia (Conseller\'\i a de Educaci\'on) and FEDER; X.F. is
supported by grant ED481B-2019-040 (Xunta de Galicia) and the Fulbright
Visiting Scholar fellowship.

\end{document}